\documentclass[pre,twocolumn,english,superscriptaddress,floatfix,longbibliography]{revtex4-2}

\usepackage{graphicx}
\usepackage{dcolumn}
\usepackage{bm}
\usepackage{bbm}
\usepackage{soul}
\usepackage{enumitem}
\usepackage{siunitx}
\usepackage{makecell}
\usepackage{physics}
\usepackage{ragged2e}
\makeatletter
\long\def\@makecaption#1#2{%
  \vskip\abovecaptionskip
  \sbox\@tempboxa{\small #1: #2}%
  \ifdim \wd\@tempboxa >\hsize
    {\small\justifying #1: #2\par}
  \else
    \global \@minipagefalse
    \hb@xt@\hsize{\hfil\box\@tempboxa\hfil}%
  \fi
  \vskip\belowcaptionskip}
\makeatother

\usepackage{braket}
\usepackage{graphicx}
\usepackage{amsmath,verbatim,latexsym,amssymb,indentfirst,mathrsfs,mathtools,amsthm,bbm,bm,url}
\usepackage{hyperref}
\hypersetup{colorlinks=true, citecolor=blue, linkcolor=black, urlcolor=blue}
\usepackage[title,titletoc]{appendix}

\renewcommand*{\ketbra}[2]{\lvert #1 \rangle\!\langle #2 \rvert}

\usepackage{cancel}
\usepackage{bbold}
\usepackage[dvipsnames]{xcolor}
\usepackage[normalem]{ulem}
\usepackage{physics}

\newcommand{\microsecond}{\mu\mathrm{s}}

\begin{document}

\title{Nonlinear quantum evolution of a dissipative superconducting qubit}
\author{Orion Lee}
\altaffiliation[These authors contributed equally to this work.]{}
\affiliation{
Department of Physics, Washington University, St. Louis, Missouri 63130, USA
}

\author{Qian Cao}
\altaffiliation[These authors contributed equally to this work.]{}
\affiliation{
Department of Physics, Washington University, St. Louis, Missouri 63130, USA
}

\author{Yogesh N. Joglekar}
\affiliation{Department of Physics, Indiana University Indianapolis, Indianapolis, Indiana 46202, USA}

\author{Kater Murch}
\email{
katermurch@berkeley.edu
}
\affiliation{
Department of Physics, Washington University, St. Louis, Missouri 63130, USA
}
\affiliation{Department of Electrical Engineering and Computer Sciences, University of California, Berkeley, Berkeley, California 94720, USA}
\affiliation{Department of Physics, University of California, Berkeley, Berkeley, California 94720, USA}

\date{\today}

\begin{abstract}
Unitary and dissipative models of quantum dynamics are linear maps on the space of states or density matrices. This linearity encodes the superposition principle, a key feature of quantum theory. However, this principle can break down in effective non-Hermitian dynamics arising from postselected quantum evolution. We theoretically characterize and experimentally investigate this breakdown in a dissipative superconducting transmon circuit. Within the circuit's three-level manifold, no-jump postselection generates an effective non-Hermitian Hamiltonian governing the excited two-level subspace and an anti-Hermitian nonlinearity. We prepare different initial states and use quantum state tomography to track their evolution under this effective, nonlinear Hamiltonian. By comparing the evolution of a superposition-state to a superposition of individually-evolved basis states, we test linearity and observe clear violations which we quantify across the exceptional-point (EP) degeneracy of the non-Hermitian Hamiltonian. We extend the analysis to density matrices, revealing a breakdown in linearity for the two-level subspace while demonstrating that linearity is preserved in the full three-level system. These results provide direct evidence of nonlinearity in non-Hermitian quantum evolution, highlighting unique features that are absent in classical non-Hermitian systems.

\end{abstract}

\maketitle

Linearity is a foundational property of quantum dynamics. Given two states $\ket{a}$, $\ket{b}$ and a Hermitian Hamiltonian $H$, the time evolution is given by $\ket{a}\to \ket{a(t)}\equiv G(t)\ket{a}$, $\ket{b}\to \ket{b(t)}\equiv G(t)\ket{b}$, where $G(t)\equiv\exp(-iHt)$ is determined by the Schr\"{o}dinger equation. Linearity implies that a state $\ket{c}=\alpha\ket{a}+\beta\ket{b}$, unitarily evolved to $\ket{c(t)}$, is equal to the linear superposition of time-evolved $\ket{a(t)},\ket{b(t)}$ states i.e. $\ket{c}\to \ket{c(t)} = \alpha \ket{a(t)}+ \beta \ket{b(t)}$~\cite{griffiths2018introduction}. The same holds for convex combinations of density matrices that evolve unitarily via the von-Neumann equation~\cite{Breuer2007}. This assumption of linearity extends to open, dissipative quantum systems. When a quantum system interacts with its environment, the resultant dynamics of its reduced density matrix $\rho(t)$ is described by the Lindblad equation $\partial_t\rho(t)=\mathcal{L}\rho(t)$~\cite{Breuer2007} that is linear in the density matrix or more generally, linear completely-positive trace-preserving maps~\cite{Nielsen2012}. Thus, linear dynamics underpin signatures of quantumness, such as spatiotemporal entanglement, and their degradation. 

In contrast, exceptional classical phenomena such as rouge waves~\cite{Chabchoub2012} result from a {\it nonlinear} Schr\"{o}dinger equation that applies to shallow fluids and optics~\cite{KEVREKIDIS2001,Peregrine1983,Fibich2015}. Here, nonlinearity encodes changes in the potential due to local wave intensity. In quantum systems, such state-dependent Hermitian nonlinearity leads to remarkable predictions~\cite{weinberg1989} including arbitrarily fast signaling~\cite{gisin1990,Polchinski1991}, closed timelike curves~\cite{Bennett2009}, and efficient solutions to classical, NP-complete problems~\cite{Abrams1998}. Experiments have constrained the dimensionless strength of such nonlinearity to less than parts per trillion~\cite{Broz2023,Polkovnikov2023} bolstering the belief that Hermitian nonlinearities are absent in nature.

In recent years, classical and quantum systems governed by an effective non-Hermitian Hamiltonian $H_\mathrm{eff}=H-i\Gamma$, have proliferated~\cite{Ashida2020}. In linear models, their dynamics do not preserve the state-norm since the corresponding time-evolution operator $G_\mathrm{eff}(t)$ is not unitary. In the classical context, this norm violation is interpreted as energy or material exchange with the environment. In quantum cases, $H_\mathrm{eff}$ arises from post-selection over no-quantum-jump trajectories~\cite{nagh19,Rotter2019}, state of the ancilla~\cite{Wu2019}, or no-photon-loss detection~\cite{Klauck2019,Maraviglia2022}. However, as the post-selected state $\ket{\psi(t)}$ is always normalized, it satisfies a nonlinear Schr\"{o}dinger equation~\cite{Brody2012} 
\begin{align}    
i\partial_t\ket{\psi(t)}=\left(H_\mathrm{eff}+i\braket{\psi|\Gamma|\psi}\right)\ket{\psi(t)}.
\end{align}
Thus, in the quantum context, an anti-Hermitian nonlinearity $\braket{\psi|H^\dagger_\mathrm{eff}-H_\mathrm{eff}|\psi}/2$ always accompanies a non-Hermitian Hamiltonian $H_\mathrm{eff}$.
These considerations motivate a systematic investigation of signatures of nonlinearity in such quantum systems. 

In this work, we investigate the breakdown of linearity in the evolution of a superconducting transmon circuit under an effective non-Hermitian Hamiltonian. We use quantum state tomography \cite{James2001, Ariano2001, Altepeter2003} to compare the trajectories arising from initial superposition states to superpositions of the trajectories arising from initial basis states.   By post-selecting on no-jump trajectories that preserve the excited two-level subspace we isolate the dynamics governed by the non-Hermitian Hamiltonian. The postselection process implies renormalization of the experimental trajectories into the postselected subset, resulting in a breakdown of linearity.   We test this by preparing basis and superposition states and compare their evolution. We demonstrate a practical tool to realize anti-Hermitian nonlinearity at the quantum level, and thereby access and control nonlinear quantum dynamics. This opens the door to new classes of quantum behavior and information processing strategies that are not available in unitary or dissipative evolution.

\begin{figure}
    \centering
\includegraphics[width=0.35\textwidth]{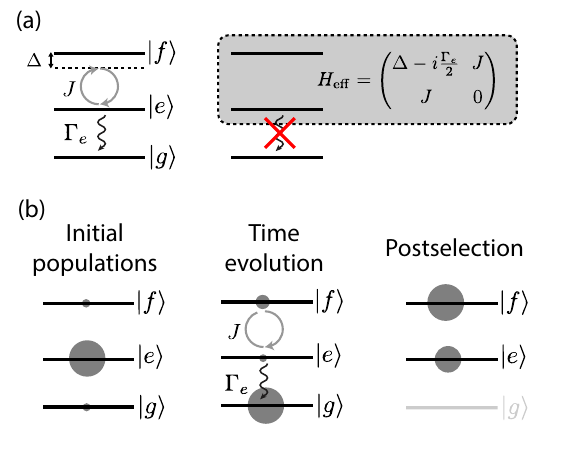}
    \caption{{\bf Effective non-Hermitian evolution via postselection on no quantum jumps.} (a) Three-level transmon manifold with states ${|g\rangle, |e\rangle, |f\rangle}$. The $|e\rangle$ state decays to $|g\rangle$ at rate $\Gamma_e$, while coherent coupling $J$ hybridizes $|e\rangle$ and $|f\rangle$ with detuning $\Delta$. (b) Illustration of non-Hermitian dynamics. Left: initial population distribution. Middle: coherent coupling and decay during time evolution. Right: postselection yielding the effective dynamics within the ${|e\rangle, |f\rangle}$ subspace.}
    \label{fig:fig0}
\end{figure}

\emph{Setup}--- To introduce how we realize effective non-Hermitian evolution with superconducting circuits, consider the three lowest levels of a transmon circuit \cite{Koch2007}, where we label the three lowest energy eigenstates as $\{\ket{g},\ket{e}, \ket{f}\}$. By engineering the transmon’s dissipative coupling to its electromagnetic environment, we realize a hierarchy of decay rates: the decay from $\ket{f}$ to $\ket{e}$, described by the dissipator $L_f = \sqrt{\Gamma_f} \ket{e}\bra{f}$, is relatively slow ($\Gamma_f = 0.057\ \microsecond^{-1}$), while the decay from $\ket{e}$ to $\ket{g}$, given by $L_e = \sqrt{\Gamma_e} \ket{g}\bra{e}$, is comparatively fast ($\Gamma_e = 0.91\ \microsecond^{-1}$). The evolution of the qutrit density matrix $\rho^{(3)}$ is governed by the Lindblad equation~\cite{Nielsen2012,Hatano2019}:
\begin{equation}
    \partial_t{\rho}^{(3)} = -i[H, \rho^{(3)}] + \sum_{j\in\{e,f\}}  \left(L_j \rho^{(3)} L_j^\dagger - \frac{1}{2} \{L_j^\dagger L_j, \rho^{(3)}\} \right).
    \label{eq:linblad}
\end{equation}
To access non-Hermitian dynamics, we consider the regime where the $\ket{f}$ decay is negligible, $\Gamma_f\ll\Gamma_e$, and quantum jumps to the $\ket{g}$ state do not occur (Fig.~\ref{fig:fig0}a). The driving Hamiltonian $H$ is given by a drive with amplitude $J$ that is detuned by $\Delta$ from the $\{\ket{e},\ket{f}\}$ transition.  Under these approximations, the evolution of the system is described by an effective non-Hermitian Hamiltonian. Expressed in the rotating frame of the $\{\ket{e}, \ket{f}\}$ basis the Hamiltonian is given by
\begin{equation}
H_{\mathrm{eff}} =
\begin{pmatrix}
\Delta - i\frac{\Gamma_e}{2} & J \\[6pt]
J & 0
\end{pmatrix}.
\label{eq:Heff_matrix}
\end{equation}
This effective Hamiltonian captures the evolution of the system conditioned on no quantum jumps and serves as the generator of dynamics for the post-selected, two-level subspace as seen in Fig.~\ref{fig:fig0}b. $H_\mathrm{eff}$ is closely related to the widely studied $\mathcal{PT}$-dimer Hamiltonian \cite{bend07, bend13mech}, which features balanced gain and loss between the two states. 
\begin{equation}
H_{\mathrm{eff}} = \frac{-i \Gamma_e}{4} \mathbb{1} + 
\begin{pmatrix}
\Delta - i\frac{\Gamma_e}{4} & J \\[6pt]
J & +i\frac{\Gamma_e}{4} 
\end{pmatrix} = \frac{-i \Gamma_e}{4} \mathbb{1} + H_\mathcal{PT}.
\label{eq:Heff_matrix}
\end{equation} 
Of particular interest is the exceptional point degeneracy, occurring at $\Delta=0$ and $J = \Gamma_e/4$, where the eigenvectors coalesce.

\begin{figure}
\includegraphics[width=0.5\textwidth]{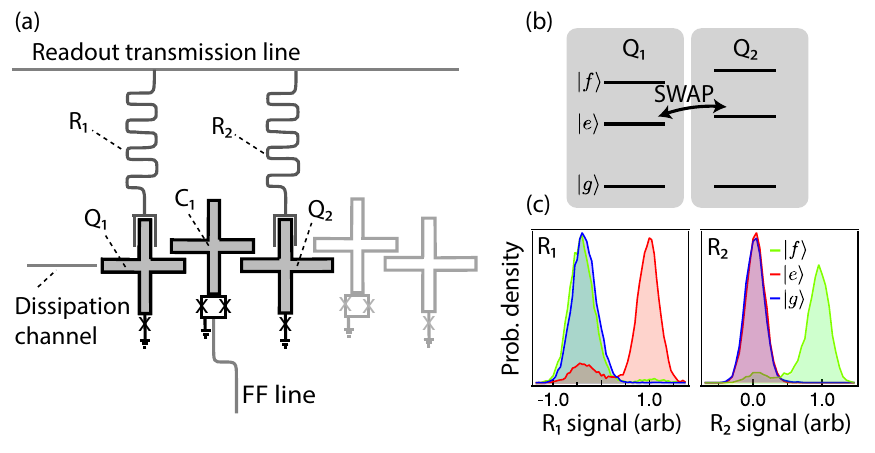}
    \caption{{\bf Setup.} (a) Schematic of the superconducting transmon processor including qubits $Q_1$ and $Q_2$, coupler $C_1$ and readout resonators $R_1$ and $R_2$. $Q_1$ is coupled to a dissipation channel leading to energy decay of the state $\ket{e}$.  (b) Parametrically activated SWAP readout: population in the $|e\rangle$ state on $Q_1$ is transferred to $Q_2$ via a parametric modulation of the coupler $C_1$. (c) Readout signal histograms display how joint readout of $R_1$ and $R_2$ yields high fidelity state assignment.}
    \label{fig:efigure1}
\end{figure}

 \emph{Quantum State Tomography}--- As depicted in Fig.~\ref{fig:efigure1}a, our experimental platform utilizes a three-transmon subsection of a multi-transmon chip.  Two transmons, $Q_1$ and $Q_2$, are dispersively coupled to dedicated microwave readout resonators, and a flux-tunable transmon, $C_1$ mediates coupling between the two. The readout resonators enable state measurement of the transmons in the energy basis. A microwave tone near resonance with the readout resonator will acquire a transmon-state-dependent phase shift, which is detected with heterodyne demodulation. A traveling-wave parametric amplifier (TWPA) is used to enhance the measurement signal-to-noise ratio \cite{mack15}. When the demodulated measurement tone is integrated for a time $t_\mathrm{meas} = 2\  \microsecond$, the signal-to-noise ratio is sufficient to distinguish between energy states with about 88\% fidelity. 

We will use the $\{\ket{g}, \ket{e}, \ket{f}\}$ manifold of states of $Q_1$  to realize the dissipative qutrit captured by Eq.~\ref{eq:linblad}. The rapid decay of $\ket{e}$ for $Q_1$ (with energy decay time $1/\Gamma_e= 1.1\ \microsecond<t_\mathrm{meas}$) would severely limit the measurement fidelity of this state. To overcome this, we implement a SWAP operation between $Q_1$ and $Q_2$, transferring population to a transmon with slower energy decay (Fig.~\ref{fig:efigure1}b). Specifically, the SWAP is implemented by parametric modulation of $C_1$'s frequency \cite{Sete2021,Yan2018}.   When the modulation frequency is equal to half the detuning between the $\ket{e}_{Q_1}\otimes \ket{g}_{Q_2}\equiv \ket{eg}$ and $\ket{ge}$ states, it brings the two states into parametric resonance. This parametric resonance results in a SWAP gate time $t_\mathrm{SWAP} = \SI{156}{\nano\second}$. Because $t_\mathrm{SWAP}\ll 1/\Gamma_e$ population is  transferred with reasonable fidelity.  Following the SWAP, we perform simultaneous readout on both qubits. 

We calibrate the readout by alternately preparing the states $\ket{g}$, $\ket{e}$, and $\ket{f}$ and perform the SWAP joint readout. Figure \ref{fig:efigure1}c displays histograms of the demodulated measurement signal for the two qubits, highlighting how the three states of $Q_1$ are  resolved with high fidelity.  We use a gradient boosting classifier \cite{gradientboosting2017} to assign measurement outcomes to each of the three logical states based on the joint readout data. The classifier produces the state assignment matrix and we use an iterative Bayesian-update correction to compensate for the measurement infidelities \cite{Nachman2020}.

\begin{figure}
    \includegraphics[width = 0.48 \textwidth]{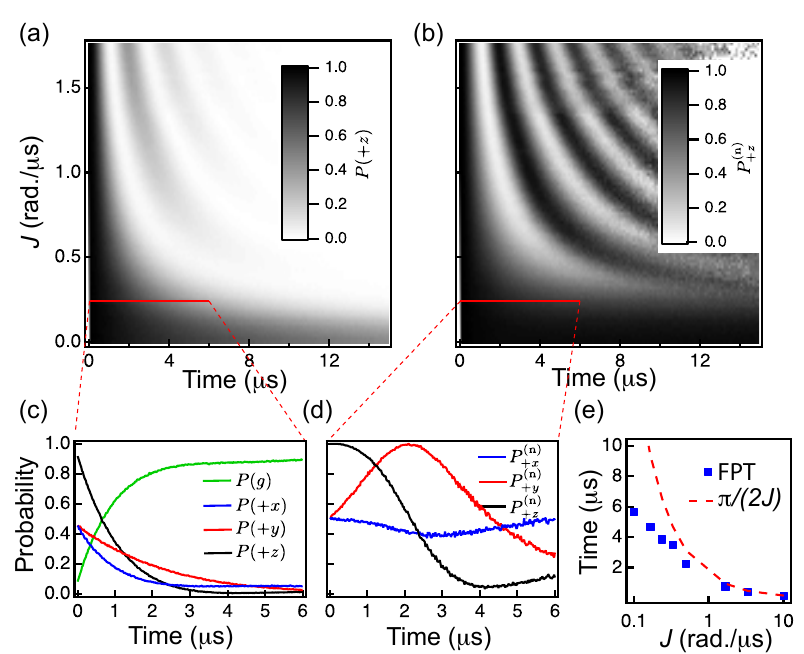}
    \caption{Characterization of the system dynamics before and after postselection. (a) $P(+z)$ versus $J$ and time. (b) The normalized $P^{\mathrm{(n)}}_{+z}$ after postselection. (c) Measurement probabilities from tomographic projections along different axes in the $\{\ket{e}, \ket{f}\}$ manifold along with the ground state population. (d) Postselected data from (c). (e) First passage time (FPT) versus $J$ (blue squares) compared to the expected FPT for Hermitian evolution ($\pi/(2J)$, dashed line). }
    \label{fig:tomo}
\end{figure}

We now focus on the dynamics within the $\{\ket{e},\ket{f}\}$ manifold in the absence of quantum jumps to $\ket{g}$.  By only keeping trajectories that preserve this manifold of states, our postselected ensemble of data is effectively renormalized.  We define Pauli operators on the $\{\ket{e},\ket{f}\}$ manifold  as $\sigma_z \equiv \ket{e}\bra{e} - \ket{f}\bra{f}$,   $\sigma_x \equiv \ket{e}\bra{f} + \ket{f}\bra{e}$, and $\sigma_y \equiv -i\ket{e}\bra{f} + i \ket{f} \bra{e}$. We measure these operators' expectation values with quantum state tomography.  The $x$-axis tomography is performed as follows. We apply a $\pi/2$ rotation in the $\{\ket{e},\ket{f}\}$ manifold followed by the SWAP based three state readout.  For each experimental trials, the classifier yields measurement outcomes ``$g$'', ``$e$'', or ``$f$'', which given the $\pi/2$ rotation correspond to measurement results ``$g$'', ``$+x$'', or ``$-x$''. By performing several trails we obtain raw probabilities $\tilde{P}(g),\ \tilde{P}(+x),\ \tilde{P}(-x)$. We then correct these probabilities using the iterative Bayesian update, yielding corrected probabilities, $P(g),\ P(+x),\ P(-x)$, finally we look at the renormalized sub-ensemble: $P^\mathrm{(n)}(+x)\equiv \frac{1}{P(+x)+P(-x)}P(+x),\ P^\mathrm{(n)}(-x)\equiv\frac{1}{P(+x)+P(-x)}P(-x)$]. We perform tomography about the $y$- and $z$-axes similarly. 

The corrected probabilities allow us to reconstruct the $2\times2$ density matrix in the $\{|e\rangle, |f\rangle\}$ basis as
\begin{equation}
    \rho_{ef} = \frac{1}{2}(\mathbb{I} + x \sigma_x + y \sigma_y + z \sigma_z ),
\end{equation}
where $x = 1-2 P^\mathrm{(n)}(+x)$, $y= 1-2 P^\mathrm{(n)}(+y)$, and $z= 1-2 P^\mathrm{(n)}(+z)$ are the experimentally determined expectation values of $\sigma_x$, $\sigma_y$, and $\sigma_z$.

Figure~\ref{fig:tomo} displays the characterization of the system dynamics with quantum state tomography. We prepare $Q_1$ in the initial state $\ket{e}$ and apply microwave driving for varying durations followed by tomography. We repeat such experiments for different values of $J$ with $\Delta\simeq 0$ fixed. Figure~\ref{fig:tomo}a displays the corrected probability $P(+z)$ versus time and $J$. Postselection on the no-jump evolution yields the renormalized sub-ensemble probabilities; $P^{(\mathrm{n})}_{+z}$  is shown in Fig.~\ref{fig:tomo}b.  The plot of $P^{(\mathrm{n})}_{+z}$ versus time and $J$ reveals two clear regions with different dynamics: for large $J$ the dynamics are oscillatory, for smaller $J$ the oscillations abruptly stop. This transition corresponds to the $\mathcal{PT}$-symmetry breaking transition, with $J>\Gamma/4$ corresponding to the $\mathcal{PT}$-symmetry unbroken region and, $J<\Gamma/4$ the $\mathcal{PT}$-symmetry broken regime. $J=\Gamma/4$ marks the exceptional point (EP) degeneracy. Figure~\ref{fig:tomo}c,d display the tomography components versus time for a selected region (red line in panels a,b) near the EP.

 Figure~\ref{fig:tomo}d highlights how dynamics under an effective non-Hermitian Hamiltonian can accelerate quantum dynamics near the EP. Here, the drive strength is $J= 0.24 \ \mathrm{rad.}/\microsecond$. In the Hermitian limit, the first passage time (FPT) (or $\pi$-pulse time) would be $T_\pi = \pi/J \simeq 13 \ \microsecond$. Instead, the dynamics show oscillation from $\ket{e}$ to $\ket{f}$ in $T_\pi \approx 4\ \microsecond$. This accelerated dynamics occurs because the dissipation and drive conspire to drive the system rapidly from $\ket{e}$ to $\ket{f}$ \cite{bender07faster}.  Figure~\ref{fig:tomo}e further explores the first passage time for different drive strengths; we can see clear deviation from the expected value as $J$ is reduced.

\emph{Testing Linearity of non-Hermitian Quantum Evolution}--- 
We now turn to testing if the non-Hermitian evolution realized through postselection on no quantum jumps preserves linearity of quantum evolution. As sketched in the introduction, our test will be based comparing the time evolution of a superposition of states a superposition of the individual states' time evolution. This notion of linearity applies to state kets but not density matrices (e.g.\ $\ketbra{+x}{+x} \neq \alpha \ketbra{e}{e} + \beta \ketbra{f}{f} \quad \forall \{ \alpha, \beta\}$).  We will revisit a special case of classical mixtures in a later section, but for now our first step is to construct pure state kets from the measured density operators.

\begin{figure}
    \includegraphics[width = 0.48\textwidth]{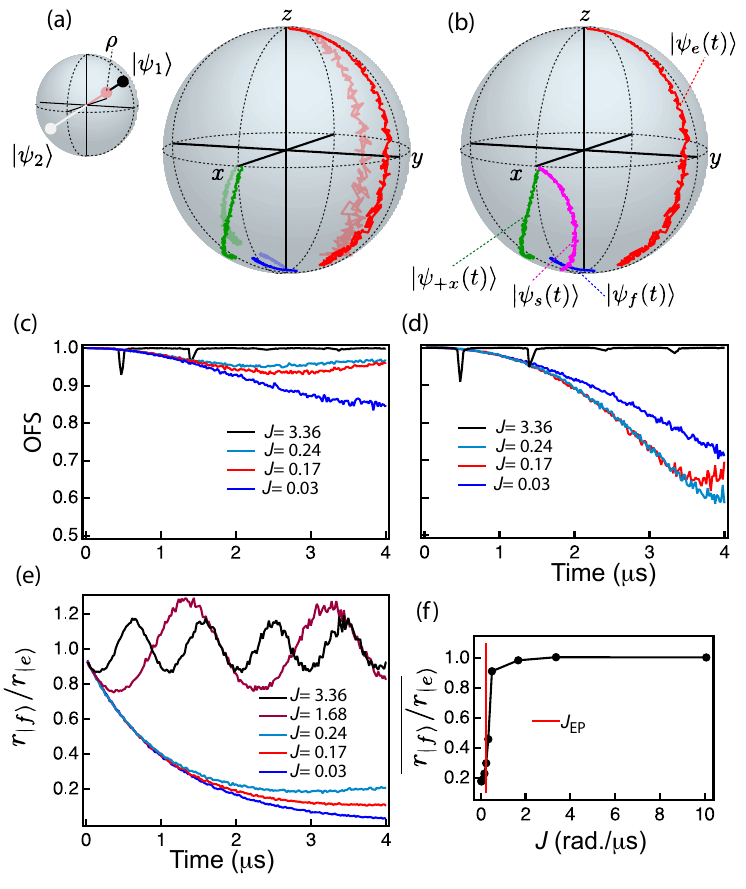}
    \caption{{\bf Testing linearity of non-Hermitian quantum evolution.} (a) The pure state projection of the tomographically reconstructed density matrix $\rho$ is given by its eigenvector with largest eigenvalue $\ket{\psi_1}$ (inset). The main panel displays three trajectories within the Bloch ball (transparent red, blue, green) reconstructed over time $t\in(0,6)\ \microsecond$ from different initial states (respectively $\ket{e},\ \ket{f},\ \ket{+x}$). The pure state projections are shown in solid colors. (b) We compare the trajectory originating in $\ket{+x}$, denoted $\ket{\psi_{+x}(t)}$ to the trajectory reconstructed from a superposition of $\ket{\psi_{e}(t)}$ and $\ket{\psi_{f}(t)}$ trajectories, denoted $\ket{\psi_s(t)}$ and displayed in magenta. (c) OFS for versus time for different values of $J$ and initial state $\ket{+x}$. (d) OFS versus time for initial state $\ket{+y}$.  (e) The postselection ratios as a function of time showing oscillations for values above the exceptional point and decay for those below. (f) The time averaged value of the postselection ratio. }
    \label{fig:purestate}
\end{figure}

Figure~\ref{fig:purestate}a displays the procedure; given a state $\rho$ within the Bloch ball, take the eigenvectors of $\rho$: $\{\ket{\psi_1}, \ \ket{\psi_2}\}$ and choose the eigenvector with greatest eigenvalue as the pure state projection. In Fig.~\ref{fig:purestate}a we display the measured trajectories within the Bloch ball and compare to the pure state projections to the Bloch sphere. We consider trajectories originating from different initial states $\ket{e}$ (red), $\ket{f}$ (blue), and $\ket{+x}$ (green).  The different dynamics for these three states again highlight the non-unitary evolution generated by $H_\mathrm{eff}$; the trajectory originating from $\ket{e}$ evolves along the $y$--$z$ longitude from $\ket{e}$ to $\ket{f}$, while the trajectory originating in $\ket{f}$ barely deviates from its initial state.  

To assess whether the measured dynamics preserve linearity, we compare the trajectory originating in $\ket{+x}$, denoted $\ket{\psi_{+x}(t)}$, to a trajectory constructed from the superposition $\ket{\psi_s(t)}\equiv A_s(\ket{\psi_{e}(t)}+\ket{\psi_{f}(t)})/\sqrt{2}$. Here $A_s$ is a normalization factor to account for non-orthogonality of $\ket{\psi_{e}(t)}$ and $\ket{\psi_{f}(t)}$. The trajectories are compared in Fig.~\ref{fig:purestate}b; the trajectory for $\ket{\psi_s(t)}$ is displayed in magenta. Clearly $\ket{\psi_{+x}(t)}\neq \ket{\psi_s(t)}$. We choose to quantify the deviation from linearity in terms of Fubini-Study-metric \cite{Provost1980} (OFS) as a function of time given by 
\begin{equation}
    \mathrm{OFS}(\theta, t) = \frac{\langle\Phi_\theta (t) | \psi_\theta(t)\rangle}{\sqrt{\langle\phi_\theta (t) | \phi_\theta(t)\rangle}}.
    \label{eqs::OFS}
\end{equation}
We expect when more linearity is preserved, this value is closer to 1. Figure~\ref{fig:purestate}c,d display the time evolution of the OFS for the initial state $\ket{+x}$ and $\ket{+y}$ respectively for different values of $J$. In both cases, large values of $J$ yield an OFS $\approx 1$ (the small deviations from 1 arise from imperfections in the state purification process). For   $J<J_\mathrm{EP}$ both cases show $\mathrm{OFS}<1$; the deviation from linearity is more pronounced in Fig.~\ref{fig:purestate}d, where the initial state $\ket{+y}$ is orthogonal to the lone eigenstate at the exceptional point; $\ket{-y}$.

The breakdown of linearity in our system arises from the requirement that quantum states remain normalized under non-Hermitian evolution. A non-Hermitian Hamiltonian by itself produces non-unitary dynamics, but this evolution remains linear. Nonlinearity emerges because the non-norm-preserving evolution generated by $H_\mathrm{eff}$ causes different initial states to accumulate population in $\ket{g}$ at different rates. This leads to trajectory-dependent postselection success probabilities and corresponding trajectory-dependent renormalization factors that break the superposition principle. We define the renormalization factor $r_{\ket{i}}(t) = \frac{1}{1-P(g)}$ based on the postselection success probability ($1-P(g)$) for initialization in state $\ket{i}$. Figure~\ref{fig:purestate}e displays the ratio $r_{\ket{f}}/r_{\ket{e}}$ versus time for different values of $J$. In the $\mathcal{PT}$-unbroken regime ($J > J_\mathrm{EP}$), this ratio oscillates around unity, indicating that the two trajectories spend comparable time in the lossy $\ket{e}$ state and receive similar renormalization. In the $\mathcal{PT}$-broken regime ($J < J_\mathrm{EP}$), the ratio decays, reflecting asymmetric population dynamics. The time-averaged ratio (Fig.~\ref{fig:purestate}f) approaches unity as $J$ increases well beyond $J_\mathrm{EP}$, where the trajectories become increasingly symmetric. This demonstrates that the exceptional point roughly demarcates the boundary between predominantly nonlinear and increasingly linear behavior over long timescales.

\begin{figure}
    \centering
    \includegraphics[width = 0.48\textwidth]{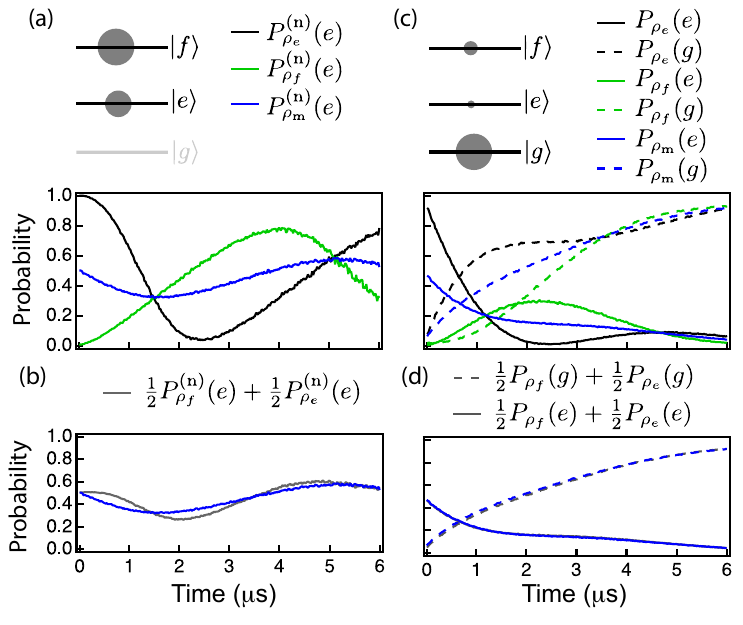}
    \caption{{\bf Linearity of classical mixtures.} 
    (a) We show the postselected probability data versus time  for different initial states; $\rho_e$, $\rho_f$ and $\rho_m$, evolved with $J = 0.5 \ \mathrm{rad.}/\microsecond$. (b) We compare $P_{\rho_\mathrm{m}}^{(\mathrm{n})}(e)$ to the classical superposition $\frac{1}{2} P_{\rho_f}^{(\mathrm{n})}(e)+\frac{1}{2}P_{\rho_e}^{(\mathrm{n})}(e)$. The curves are in disagreement arising from the breakdown of linearity for classical mixtures.  (c) We display the probabilities in the un-postselected three state system for the same initial preparations. (d) Comparing the evolution from the $\rho_\mathrm{m}$ preparation to the classical superposition shows excellent agreement, verifying linearity for the three-state dissipative qubit.}
    \label{fig:classical}
\end{figure}

Having studied the breakdown of linear quantum evolution for superpositions of kets, we now turn to a special case: superpositions of classical mixtures. We previously noted that the superposition principle does not extend in general to density operators. However, it can be expressed in terms of classical mixtures (i.e.\ with no coherence). This case is particularly interesting because it allows us to investigate linearity in both the postselectected non-Hermitian qubit as well as for the dissipative three level system spanned by $\{\ket{g}, \ket{e}, \ket{f}\}$.  

As shown in Fig.~\ref{fig:classical}(a) we first display the time evolution of populations in the $\{\ket{e}, \ket{f}\}$ manifold, simply displayed as $P^{(\mathrm{n})}(e)$. We consider three different initial states: $\rho_e = \ket{e}\bra{e}$ (with $P(e) = 1$), $\rho_f$ (with $P(e)=0$), and a classical mixture $\rho_\mathrm{m}$ with $P(e) = 0.5$. The mixture is composed of equal parts preparations in $\ket{+x}$ and $\ket{-x}$. The dynamics for the populations are characteristic of the non-Hermitian evolution in this manifold with asymmetric oscillation favoring population in $\ket{f}$ over $\ket{e}$. The lower panel compares the evolution of the initial mixture to a superposition of the curves for the $\ket{e}$ and $\ket{f}$ preparations. While the curves are close, there is clear disagreement---a breakdown of linearity for classical mixtures. 

Figure~\ref{fig:classical}(b) extends this analysis to the three state system. Here we retain all measurement outcomes after the iterative Bayesian update, $P(g)$, $P(e)$, and $P(f)$ which can be arranged in a diagonal density matrix, 
\begin{equation}
    \rho^{(3)} =
    \begin{pmatrix}
        P(g) & 0 & 0 \\
        0 & P(e) & 0 \\
        0 & 0 & P(f)
    \end{pmatrix}.
\end{equation}
Figure~\ref{fig:classical}(b) displays $P(e)$ and $P(g)$ for the same three preparations. The lower panel tests if the superposition of the density matrices preserves linearity: it does.

\emph{Outlook}--- 
Our experimental demonstration of nonlinearity in postselected quantum evolution represents a departure from both Bender's original formulation of $\mathcal{PT}$-symmetric quantum theory \cite{bend98}, which sought to preserve unitarity through modified inner products, and subsequent classical realizations of non-Hermitian dynamics in photonic and mechanical systems. While the mathematical isomorphism between classical dynamical matrices and quantum Hamiltonians has enabled extensive exploration of non-Hermitian phenomena across diverse platforms, the quantum implementation through postselection on no-jump trajectories introduces a fundamentally new element: state-dependent renormalization that breaks the superposition principle. This nonlinearity, absent in classical non-Hermitian systems and unitary quantum evolution alike, emerges as a uniquely quantum feature of postselected dynamics. Our characterization of where nonlinear regimes become prominent---particularly near exceptional points but extending throughout the parameter space---elucidates the trade-offs inherent in accessing these dynamics. While nonlinear quantum evolution has been theoretically predicted to offer computational advantages for certain problems, our results highlight that these benefits come at the cost of reduced postselection success rates and probabilistic operation. Future work exploring applications of controlled nonlinearity in quantum information processing must therefore balance the potential for enhanced computational power against the practical limitations imposed by postselection overhead. The ability to systematically tune and characterize this nonlinearity in a well-controlled superconducting platform opens new avenues for investigating fundamental questions about the boundaries of quantum mechanics and developing novel quantum technologies that leverage the unique features of postselected evolution.

\begin{acknowledgments}
Devices were fabricated and provided by the Superconducting Qubits at Lincoln Laboratory (SQUILL) Foundry at MIT Lincoln Laboratory, with funding from the Laboratory for Physical Sciences (LPS) Qubit Collaboratory.
This work received support from the National Science Foundation award No.~PHY-2408932, the Air Force Office of Scientific Research (AFOSR) Multidisciplinary University Research Initiative (MURI) Award on Programmable systems with non-Hermitian quantum dynamics (Grant No. FA9550-21-1-0202), and ONR Grant No. N000142512160.
\end{acknowledgments}


\pagebreak
\begin{widetext}

\newpage

\section*{Supplementary Information}

The Supplementary Information  contains details about the experimental setup and qubit chip, information about the multi-state readout. 

\subsection{Setup}

The experimental setup comprises a multiqubit superconducting chip fabricated and provided by the Superconducting Qubits at Lincoln Laboratory (SQUILL) Foundry at MIT Lincoln Laboratory. The device is packaged, shielded, and mounted in a dilution refrigerator in a setup similar to \cite{song25}. The fast flux lines for the couplers are filtered with 30 dB attenuation and 1300~MHz low-pass filters (MiniCircuits 1300 VLFX). We use the readout input line to drive the qubit and send readout pulses. The input readout line has total 60 dB attenuation, a 8~GHz low-pass filter and dissipative filters using eccosorb epoxy. For the readout output line, we use a traveling-wave parametric amplifier (TWPA) and a high-electron-mobility transistor (HEMT) to amplify the output signal. The pump frequency of TWPA is $5.1\ \mathrm{GHz}$. 

The device features 5 qubits and four tunable couplers. This work utilizes two of those qubits with resonant frequecies $4.484$~GHz and $4.445$~GHz. Each qubit is coupled to a microwave readout resonator with respective frequencies $6.727$~GHz and $6.655$~GHz.

\subsection{Three state readout}
The SWAP-assisted three state readout relies on simultaneous measurements of both $Q_1$ and $Q_2$. We use a heterodyne method to demodulate the readout signals from two readout resonators. A single measurement yields a 4-tuple that needs to be classified into a single state. The classifier is from the LightGBM package \cite{gradientboosting2017}.

The success of the SWAP-assisted readout and classifier can be assessed by the normalized three-state classification (confusion matrix). 
\begin{equation}
\boldsymbol{\beta} =
\begin{pmatrix}
0.993 & 0.003 & 0.005 \\
0.123 & 0.871 & 0.006 \\
0.056 & 0.018 & 0.925
\end{pmatrix},
\label{eq:beta_this_work}
\end{equation}
where $\beta_{ij}$ is the probability of assigning the qubit to the $j$th state after preparing it in the $i$th state. We use an iterative Bayesian update method to compensate for the measurement infidelities \cite{Nachman2020}.

\end{widetext}
\end{document}